\documentclass[%
 reprint,
 amsmath,amssymb,
 aps,
 prl,
]{revtex4-1}

\usepackage{graphicx}
\usepackage{dcolumn}% Align table columns on decimal point
\usepackage{bm}% bold math
\begin{document}

\preprint{APS/123-QED}

\title{Chirality-induced helical self-propulsion of cholesteric liquid crystal droplets}% Force line breaks with \\

\author{Takaki Yamamoto}
 \email{tyamamoto@daisy.phys.s.u-tokyo.ac.jp}
\author{Masaki Sano}%
\affiliation{%
Department of Physics, The University of Tokyo, 7-3-1 Hongo, Bunkyo-ku, Tokyo 113-0033, Japan
}%

\date{\today}% It is always \today, today,

\begin{abstract}
We report the first experimental realization of a chiral artificial microswimmer exhibiting the helical motion.
We found that a cholesteric liquid crystal droplet with a helical director field swims in a helical path driven by the Marangoni flow in an aqueous surfactant solution. 
We confirmed that the handedness of the droplet determines that of the helical path.
This result strongly suggests that the helical motion is originated from the chirality of the cholesteric liquid crystal.
To study the mechanism of the emergence of the helical motion, we propose a coupled time-evolution equations in terms of a velocity, an angular velocity and a tensor variable representing the symmetry of the helical director field of the droplet. 
Our model shows that the chiral coupling terms between the velocity and the angular velocity play a crucial role in the emergence of the helical swimming of the droplet.

\end{abstract}

\maketitle
The role of the chirality in microswimmers is receiving increasing attention in the field of biological physics\cite{julicher_ca2,chiral_bio1,helical1}.
Owing to the biomolecular chirality, chiral motions are remarkable in biological microswimmers;
for instance, a sperm of a sea urchin controls its flagellum to steer in a circular and a helical path\cite{sperm2, sperm1}.
Recently, to study the microswimmers' dynamics without difficulties in controlling biological systems, experimental setups of artificial swimmers have been established and their dynamics has been intensively investigated\cite{dreyfus,maas}.
K\"{u}mmel et al. experimentally observed the circular motion of a chiral artificial swimmer in two dimensions (2D) using a L-shaped self-propelled particle\cite{lowen1}.
However, there have been no studies reporting a helical motion of an artificial swimmer in 3D yet.

In this Letter, we present the first experimental realization of a chiral artificial microswimmer showing the helical motion and propose a theoretical model of the chiral dynamics.
In experimental study, we focused on self-propulsions of droplets in/on fluid\cite{izri1,bartolo1,marangoni1}, such as a moving alcohol droplet on the water\cite{nagai1}. 
The underlying mechanism of the self-propulsion is the Marangoni flow induced by the gradient of the surface tension at the liquid-liquid interface\cite{schmitt1}.
Nematic liquid crystal (NLC) droplets can also swim in a surfactant solution\cite{herminghaus1}.
Meanwhile, a cholesteric liquid crystal (CLC), which shows a helical director field due to the chirality of molecules, undergoes rotational motion when subjected to external fields such as a temperature gradient\cite{lehmann1,tabe2,oswald1,yamamoto1}.
This rotational motion is explained as a result of the coupling between axial (rotational motion) and polar (external field) vectorial quantities only allowed in chiral systems\cite{groot1}.
Hence, a CLC droplet in a surfactant solution can swim in a chiral path owing to the chiral coupling between the Marangoni flow and rotational motion.

In theoretical study, we apply the model of self-propulsion of the droplets without intrinsic polarity, for which the spontaneous symmetry breaking (SSB) is essential in contrast to the propulsion of polar self-propelled particles such as asymmetric colloidal particles\cite{jiang1}.
Tarama and Ohta proposed a model of such system\cite{tarama1,ohta1}, where the dynamics of the particle obeys a coupled time-evolution equations in terms of a velocity, an angular velocity and a tensor variable representing the geometry of the particle.
Introducing coupling terms between the velocity and angular velocity allowed only in chiral systems into their model, we construct a model of the chiral self-propelled droplets.
\paragraph{Experiment}
A mixture of the NLC (4-cyano-4'-pentylbiphenyl, TCI) and 2wt\% of a chiral dopant (R811, WuXi AppTec Co., Ltd), which works as a CLC, was dispersed in an aqueous surfactant solution (10wt\%, tetradecyltrimethylammonium bromide\cite{herminghaus1}, Wako) by pipetting. 
We prepared a sample putting the solution between two cover glasses with a spacer thicker than $300\ \mu \rm{m}$.
We used such spacers much thicker than diameters of CLC droplets to observe the 3D dynamics.
The dynamics of CLC droplets in the surfactant solution was observed by an inverted optical microscope without any polarizers (DMI6000B, Leica). 
\begin{figure}[t]
\includegraphics[width = 8.5cm]{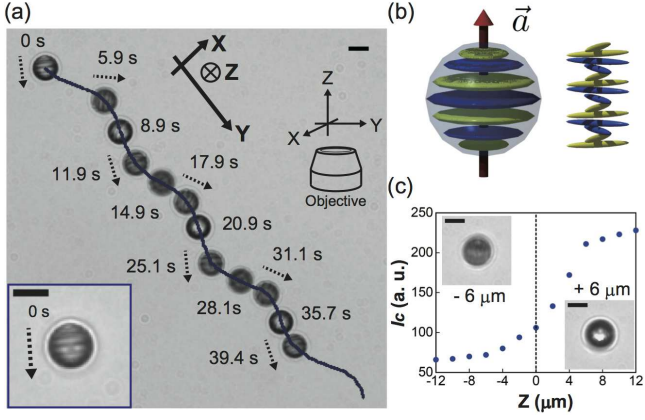}% Here is how to import EPS art
\caption{\label{fig1} (a) Time-evolution of a swimming CLC droplet ($D=13\ \mu\mbox{m}$). 
The trajectory is drawn as a solid curve. 
The dotted arrows indicate the direction of the helical axis of the droplet. 
(Inset) An enlarged image at $t=0\ \mbox{s}$. The definition of the coordinate axes is also shown. 
(b) A schematic image of a CLC droplet with a homogeneous helical director field. 
A spheroid represents a CLC molecule. 
(c) Dependence of $I_c$ on $Z$ for a droplet attached to a glass.
We take $Z=0$ at the focal plane. 
(Inset) Images of the droplet at $Z=\pm 6\ \mu\mbox{m}$.
The scale bars are all $10\ \mu\mbox{m}$.}
\end{figure}

Figure~\ref{fig1}(a) shows a time-evolution of a swimming CLC droplet (See also Supplemental Movie M1\cite{sm1}).
In the inset of Fig.~\ref{fig1}(a), the striped pattern of the droplet shows that the droplet has a homogeneous helical director field as illustrated in Fig.~\ref{fig1}(b).
Since the striped pattern is observed at each time in Fig.~\ref{fig1}(a), we find that the droplet swims without deformation of the helical director field.
Note that the droplets with the striped pattern are observed when the diameter is less than $\sim20\ \mu \rm{m}$. 
Larger droplets exhibit erratic motions with the director field disturbed by the Marangoni flow (See also Supplemental Movie M2\cite{sm1}).
In this paper, we focus on the dynamics of the droplet with the striped pattern.

\begin{figure}[t]
\includegraphics[width = 8.5cm]{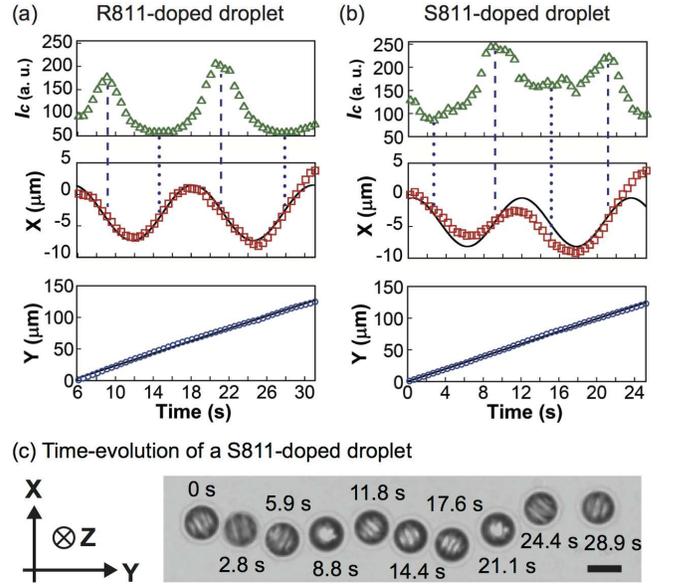}% Here is how to import EPS art
\caption{\label{fig2} Time-evolution of $(X(t), Y(t))$ and $I_c(t)$ of (a) R811-doped and (b) S811-doped droplets.
Solid curves and lines are results of sinusoidal and linear fitting, respectively.
Dotted and dashed lines indicate the local minimum and maximum of $I_c(t)$, respectively.
(c) Time-evolution of a S811-doped droplet ($D=12\ \mu\mbox{m}$). The scale bar is $10\ \mu\mbox{m}$. 
}
\end{figure}

The solid curve in Fig.~\ref{fig1}(a) shows a sinusoidal trajectory of the droplet\cite{sm3}. 
We concluded that the droplet swims in a 3D helical path for the following reasons. 
The position of the droplet at time $t$ is defined as $(X(t), Y(t), Z(t))$.
To simplify the analysis, the Y-axis of the reference frame is taken to be parallel to the axis of the 2D sinusoidal trajectory as shown in Fig.~\ref{fig1}(a).
Figure~\ref{fig2}(a) shows the time-evolution of the XY coordinate $(X(t), Y(t))$ and the averaged intensity $I_c(t)$ at the center square ($3\ \mu\mbox{m}\times3\ \mu\mbox{m}$) of the droplet. 
Here, the point at $t=5.9\ \mbox{s}$ in Fig.~\ref{fig1}(a) is taken as the origin of the XY plane.
$I_c(t)$ was calculated to investigate the dynamics in the Z-direction. 
We confirmed that $I_c$ is an increasing function of $Z$ around the focal plane $Z=0$ (Fig.~\ref{fig1}(c)). 
Here, we calculated the dependence of $I_c$ on $Z$ by capturing images of a droplet attached to a glass at each focal depth. 
In Fig.~\ref{fig2}(a), the linearity of $Y(t)$ means that the droplet moves with constant velocity $v_{\rm H}$ in the direction of the helical axis. 
The linear fitting provided $v_{\rm H}=4.9\ \mu\mbox{m}/\mbox{s}$. 
$X(t)$ is fitted well with a sinusoid $X(t)=A\sin{(\omega_{\rm f}t+\phi)} +B$, resulting in that the angular frequency $\omega_{\rm f}$ and amplitude $A$ are $0.5\ \mbox{rad}/\mbox{s}$ and $4.4\ \mu\mbox{m}$, respectively. 
Furthermore, $I_c(t)$ oscillates with the same frequency as that of $X(t)$, and the peak of $I_c$ is $\pi/2$ delayed with respect to that of $X(t)$. 
Considering the monotonicity of $I_c(Z)$ in Fig.~\ref{fig1}(c), we find that the the droplet swims in the left-handed helical path. 
Here, a helix is left-handed, if the rotational direction is anti-clockwise. 

Figure~\ref{fig2}(c) shows a trajectory of a droplet doped with S811, which is the enantiomer of R811, instead of R811. 
In Fig.~\ref{fig2}(b), we analyzed the dynamics of the S811-doped droplet in the same way as the R811-doped case, which provided $v_{\rm H}=4.9\ \mu\mbox{m}/\mbox{s}, \omega_{\rm f}=0.5\ \mbox{rad}/\mbox{s}$ and $A = 3.8\ \mu\mbox{m}$.
While the dynamics in 2D is similar to that in the R811-doped case, the peak of $I_c$ is $\pi/2$ advanced with respect to the peak of $X(t)$. 
We thus find that the S811-doped droplet exhibits the right-handed helical path, opposite to that of the R811-doped droplet. 
This result strongly suggests that the helical motion is induced by the chirality of the CLC droplet. 
In the case of the self-propelled NLC droplet, the Marangoni flow was observed inside the droplet in the translational direction\cite{herminghaus1}.
Hence, also in the CLC droplet, the Marangoni flow should occur spontaneously to determine the direction of the self-propelled motion. 
Consequently, we conclude that the helical motion is driven by the chiral coupling between the Marangoni flow and the rotational motion through the helical director field of the CLC droplet. 
Based on the above considerations, we construct a simplified model of the swimming CLC droplet by symmetry argument.
\paragraph{Theory}
We describe the dynamics of a CLC droplet by a coupled Eqs.~(\ref{eq_v})-(\ref{eq_Q}) in terms of a velocity $\bm{v}$, an angular velocity $\bm{\omega}$ and a second rank traceless symmetric tensor $Q_{ij}$ representing the helical director field inside the droplet. 
Our model does not explicitly contain the surfactant concentration field, which induces the Marangoni flow.
However, such a reduced model was recently derived in the self-propulsion of the isotropic self-propelled droplet driven by the Marangoni flow by removing the degrees of freedom of the concentration field\cite{yabunaka1,yoshinaga1}.

We define $Q_{ij} =1/2(3a_ia_j-\delta_{ij})+P/2(b_ib_j-c_ic_j)$ to describe the global symmetry of a whole CLC droplet with a helical director field.
$\bm{a},\bm{b}$ and $\bm{c}$ are unit vectors.
The primary axis $\bm{a}$ represents the direction of the helical axis of the director field (Fig.~\ref{fig1}(b)).
Note that the CLC molecules themselves align perpendicularly to $\bm{a}$. 
Hence, the meaning of $Q_{ij}$ in our model is different from that of the tensor order parameter defined in the NLC\cite{degennes,ravnik1}.
We may also consider a biaxial parameter $P\geq0$ and the secondary axes $\bm{b}$ and $\bm{c}$ ($\bm{a}\perp\bm{b}, \bm{b}\perp\bm{c}$), since, if we look at the plane crossing the center of the droplet and perpendicular to $\bm{a}$, CLC molecules align in a certain preferred direction. 
We focus on the CLC droplet with the homogeneous helical director field without deformation.
Therefore, we assume that the biaxial parameter $P$ is constant in time, while the corresponding vectors $\bm{a},\bm{b},\bm{c}$ change the direction with the angular velocity $\bm{\omega}$ as described in Eq.~(\ref{eq_Q}).

We constructed the following time-evolution equations by considering the possible terms in a chiral system and keeping some relevant terms:
\begin{eqnarray}
\nonumber \cfrac{d v_i}{dt}&=&\gamma v_i-v_jv_jv_i+a_1Q_{ij}v_j+a_2\epsilon_{ijk}\omega_jv_k\\
 &&+\mu^{\rm{iso}}\omega_i+\mu Q_{ij}\omega_j\label{eq_v}\\
\nonumber\cfrac{d\omega_i}{dt}&=&\zeta\omega_{i}-\omega_j\omega_j\omega_i+c_1Q_{ij}\omega_j+c_2\epsilon_{ijk}Q_{jl}v_l v_k\\
&&+\nu^{\rm{iso}}v_i+\nu Q_{ij}v_j\label{eq_omg}\\
\cfrac{dQ_{ij}}{dt}&=&\epsilon_{kjl}Q_{ik}\omega_l-\epsilon_{ikl}\omega_lQ_{kj} \label{eq_Q}.
\end{eqnarray}
We follow the Einstein summation convention and $\epsilon_{ijk}$ is a Levi-Civita symbol.
Note that, the coefficients, $\gamma,\zeta,a_1,a_2,c_1,c_2$, are scalars allowed even in achiral systems, while $\mu^{\rm iso},\mu,\nu^{\rm iso},\nu$ are pseudo scalars allowed only in chiral systems. 

The meaning of each term with scalar coefficients in Eqs.~(\ref{eq_v})-(\ref{eq_Q}) is as follows:
The first and second terms of the right-hand side of Eqs.~(\ref{eq_v}) and (\ref{eq_omg}) represent the self-propulsion in the translational and rotational motion as a result of the SSB, respectively. 
If $\gamma$ or $\zeta$ are positive, the droplet has an injection of the energy, resulting in the self-propelled dynamics\cite{ohta1,tarama1}; otherwise, the terms mean a damping force and torque.
We consider that $\gamma>0,\zeta<0$ in our experiments, since the dynamics of the droplet is triggered by the gradient of the surface tension, which results in the Marangoni flow and then translational self-propulsion.
In general, the rotational self-propulsion may exist in some systems. 
For instance, a water droplet on the silicon oil rotates spontaneously when vibrated vertically\cite{ebata1}.
In such a case, $\zeta$ should be positive.
The terms with the coefficients $a_1$ and $c_1$ represent the anisotropy of self-propulsion or damping.
The signs of the coefficients determine the easy axis of the translation or rotation.
If the coefficients are positive, $\bm{a}$ is the easy axis. 
Otherwise, the secondary axis $\bm{c}$ is the preferred direction.
The terms with $a_2$ and $c_2$ are the lowest order terms describing the achiral coupling between $\bm{v}$ and $\bm{\omega}$ in each time-evolution equation.
The term with $a_2$ means the turning of the translational direction due to the rotation. 
$a_2$ should approach $1$ from $0$, as the coupling between $\bm{v}$ and $\bm{\omega}$ gets stronger.
We discuss the term with $c_2$ in detail later.

The most essential part of our model is the coupling terms with pseudo scalar coefficients $\mu^{\rm iso},\mu,\nu^{\rm iso},\nu$, which represent the chiral coupling between the translation induced by the Marangoni flow and the rotation through the helical director field of the CLC droplet.
Since these terms are the off-diagonal couplings in the linear non-equilibrium thermodynamics\cite{groot1}, we expect the reciprocal relation and hence assume that $\mu^{\rm iso}=\nu^{\rm iso}$ and $\mu=\nu$. 

Numerical simulations were performed based on Eqs.~(\ref{eq_v})-(\ref{eq_Q}) by the 4th-order Runge-Kutta method ($\Delta t=1.0\times10^{-4}$).
We investigate the dynamics by changing the strength of the self-propulsion $\gamma$, the chiral couplings $\mu^{\rm iso},\mu,\nu^{\rm iso},\nu$ and $c_2$.
For simplicity, we control the strength of chiral couplings with a pseudo scalar parameter $\mu$ by setting $\mu=\nu, \mu^{\rm iso}=\nu^{\rm iso}=0.7\mu$. 
The other parameters are fixed as $a_1=-1, a_2=0.9, \zeta=-3, c_1 =0$.
Here, $a_1$ is set negative for the following reason. 
It is reported that the effective viscosity in the direction of the helical axis is comparably higher than that perpendicular to the helical axis\cite{chandra1}.
Hence, the Marangoni flow, which induces the translation, is likely to occur perpendicularly to the helical axis to minimize the dissipation. 
$a_2$ is set to be smaller than 1, since the surfactant concentration field, which is the origin of the translation $\bm{v}$, should not fully follow the rotation $\bm{\omega}$ due to the diffusivity. 
In addition, we consider isotropic damping of rotational motion ($\zeta<0, c_1=0$), since the droplet is spherical.
Furthermore, we examine the effect of the biaxiality by setting $P=0$ or $P=0.1$. 
\begin{figure}[h]
\includegraphics[width = 8.5cm]{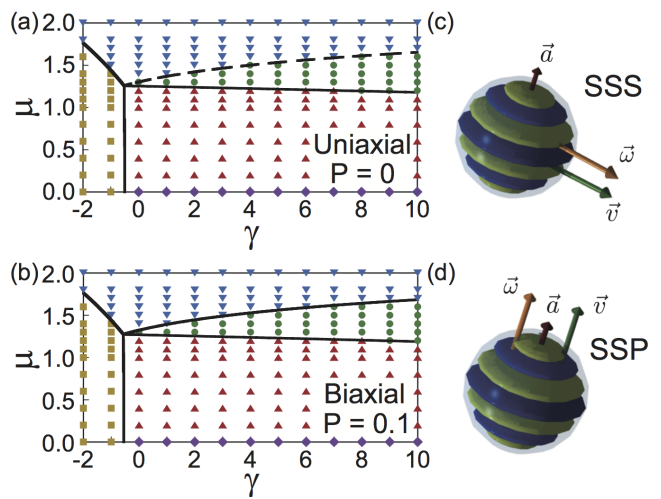}% Here is how to import EPS art
\caption{\label{fig3} (a)(b) Phase diagrams in the uniaxial ($P=0$) and biaxial case ($P=0.1$), respectively (square: NM, upward triangle: SSS, downward triangle: SSP, diamond: ST and circle: H). 
The lines indicate linear stability limit (Solid line: pitchfork bifurcation, dashed line: Hopf bifurcation).
(c)(d) Schematic images of SSS and SSP, respectively. 
Each arrow represents $\bm{a},\bm{v}$ and $\bm{\omega}$.}
\end{figure}

Figure~\ref{fig3}(a) shows the phase diagram as results of numerical simulations and linear stability analysis of Eqs.~(\ref{eq_v})-(\ref{eq_Q}) in both the uniaxial and biaxial case ($c_2=0.4$).
We numerically obtained five phases: No motion (N), Straight motion (ST), Spinning Straight motion (SS) along Secondary axis (SSS) and Primary axis (SSP), and Helical motion (H).
In N phase, the droplet is motionless.
In ST phase, the droplet moves perpendicularly to $\bm{a}$ without any rotation, which is observed in the limit of $\mu=0$.
SS is a phase where the droplet moves with $\bm{v}\parallel\bm{\omega}$. 
SS can be classified into two phases, SSS and SSP, where $\bm{v}$ is perpendicular and parallel to $\bm{a}$, respectively (See Fig.~\ref{fig3}(c)(d)).
Since SSS and SSP are predicted for the first time in our model, we hope that the two phases are experimentally realized.
When $\bm{v}\nparallel\bm{\omega}$, we classify the motion into H phase.
In H phase, the helix was right-handed, when $\mu>0$.
Importantly, changing the sign of $\mu$, namely, inversion of the chirality, provided a mirror image of the dynamics, consistent with our experiments.

In Fig.~\ref{fig3}(a)(b), the lines indicate the linear stability limit of NM, SSS and SSP phases\cite{sm3}, consistent with the numerical results.
In the uniaxial case $P=0$, the pitchfork bifurcations occur at the N--SSS, N--SSP and SSS--H boundaries, whereas the Hopf bifurcation occurs at the SSP--H boundary.
The Hopf bifurcation at the SSP--H boundary in the uniaxial case is probably related to the rotational symmetry around $\bm{a}\parallel \bm{v}\parallel \bm{\omega}$ in SSP.
In contrast, this rotational symmetry is broken in the biaxial case. 
As a result, we obtained the pitchfork bifurcation at all the boundaries in the biaxial case $P=0.1$.

\begin{figure}[t]
\includegraphics[width = 8.5cm]{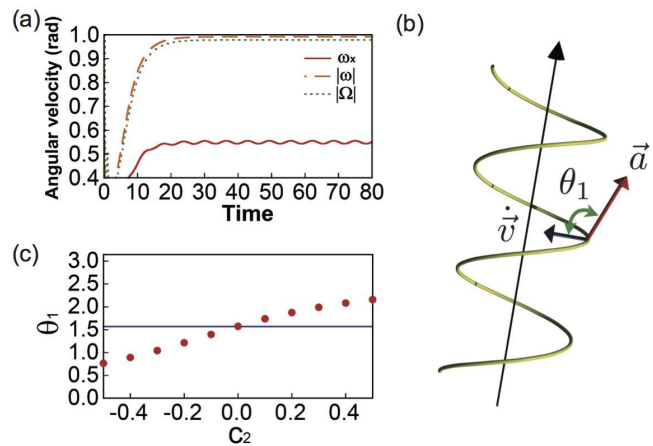}% Here is how to import EPS art
\caption{\label{fig4} (a) Time-evolution of $x$ component $\omega_x$ of $\bm{\omega}$, absolute values of $\bm{\omega}$ and $\bm{\Omega}$  in a helical motion ($P=0, c_2=0.4, \gamma=6, \mu=1.4$). 
(b) Schematic image of the definition of $\theta_1$, which is an angle between $\bm{a}$ and $\dot{\bm{v}}.$ The helical path is drawn as a helix. 
(c) Relation between $\theta_1$ and $c_2$ in the uniaxial case ($\gamma=6, \mu=1.4$). 
The solid line represents $\theta_1=\pi/2$. }
\end{figure}

In our model, whether the droplet is uniaxial or biaxial also plays a key role in the dynamics of helical motion.
In both $P=0$ and $P=0.1$, we numerically found that the helical path is the perfect helix, that is, the curvature $\kappa$ and torsion $\tau$ of the path are time-independent in the steady state.
$|\bm{v}|$ and $|\bm{\omega}|$ are also time-independent.
In contrast, the detail of rotational motion is different in both cases.
We define the angular velocity $\bm{\Omega}$ of the helical motion, which is parallel to the helical axis of the path.
In the biaxial case $P=0.1$, $\bm{\omega}$ coincides with $\bm{\Omega}$.
In the uniaxial case, oscillation of the components of $\bm{\omega}$ and a deviation between $|\bm{\omega}|$ and $|\bm{\Omega}|$ are observed (See Fig.~\ref{fig4}(a)).
The oscillation and deviation should be originated from a limit cycle in H phase, consistent with the Hopf bifurcation in the uniaxial case.
The insight in the uniaxial limit $P=0$ will be important, since the biaxiality of the CLC droplet should decrease to vanish as the wave number $q$ of the helical director field gets larger.

Finally, we discuss the term with $c_2$.
As the dotted arrows in Fig.~\ref{fig1}(a) indicate, we find a feature that the helical axis of the droplet is directed toward outside from the helical axis of the helical path.
To quantify this feature, we define $\theta_1= \arccos{(\dot{\bm{v}}\cdot\bm{a}/|\dot{\bm{v}}|)}$ (See Fig.~\ref{fig4}(b)).
Here, $\dot{\bm{v}}$ is perpendicular to the helical axis of a perfect helical path.
Hence, when $\bm{a}$ is directed outside like the experimental result, $\theta_1>\pi/2$.
The effect of $c_2$ on $\theta_1$ is numerically investigated in Fig.~\ref{fig4}(c), which suggests that $\theta_1$ is larger than $\pi/2$ when $c_2>0$. 
Accordingly, our model predicts that $c_2>0$ in the CLC droplet experiment.
Meanwhile, the second-rank symmetric tensor $v_iv_j$ in the term with $c_2$ represents the symmetry of the flow field generated by the force dipole, with which the droplet is classified into pusher or puller in the squirmer model\cite{squirmer1,squirmer2}.
This term is probably derived from the torque on the helical director field due to a force dipole. 
Although further investigation on the flow field is essential to understand $c_2$, our model suggests the importance of the force dipole in dynamics of the CLC droplet.

We have reported the chirality-induced helical motion of the self-propelled droplets for the first time both in the experiment and theory.
Although there have been a few theoretical models describing the helical motion of the self-propelled droplets\cite{tarama1,hiraiwa1}, the left- and right-handed helices appear with equal probability in such models because the chiral symmetry is not broken in the models.
Our next step will be the experimental test of the bifurcation behavior predicted in our model.
The model parameters $\gamma$ and $\mu$ are expected experimentally controllable by changing concentrations of the surfactant and chiral dopant, respectively.
To experimentally distinguish between SSS and SSP phases, we need to develop the real-time 3D tracking technique to observe the time-evolution of the director field of the droplet more precisely.

\begin{acknowledgements}
We thank H. R. Brand, T. Hiraiwa, K. H. Nagai, T. Ohta and M. Tarama for fruitful discussions. 
We thank T. Hiraiwa, K. Kawaguchi and K. H. Nagai for careful reading and essential comments on our manuscript.
This work is supported by Grant-in-Aid for JSPS Fellows (Grant No. 269814), and MEXT KAKENHI Grant No. 25103004.
\end{acknowledgements}

%\bibliography{citation}% Produces the bibliography via BibTeX.
\providecommand{\noopsort}[1]{}\providecommand{\singleletter}[1]{#1}%

\end{document}